\documentclass[twocolumn,pra,psfig,showpacs,superscriptaddress]{revtex4}
\usepackage{amsfonts}
\usepackage{graphicx}
\usepackage{amsmath}
\usepackage{amssymb}

\begin{document}

\title{Realization of the Noncommutative
Seiberg-Witten Gauge Theory by Fields in Phase Space%
\thanks{%
E-mail: ronni@gmail.com, fkhanna@ualberta.ca, adolfo@cbpf.br,
jmalboui@ufba.br, a.berti.santana@gmail.com}}
\author{R.G.G. Amorim}
\affiliation{Faculdade Gama,Universidade de Bras\'{\i}lia, 72444-240,
Bras\'{\i}lia, DF, Brazil}
\affiliation{International Centre for Condensed
Matter Physics Instituto de F\'{\i}sica, Universidade de Bras\'{\i}lia,
70910-900, Bras\'{\i}lia, DF, Brazil}
\author{F.C. Khanna}
 \affiliation{Department of Physics
and Astronomy, University of Victoria, 3800 Finnerty Road Victoria,BC V8P 5C2}
\affiliation{TRIUMF, 4004, Westbrook mall, Vancouver, British Columbia V6T 2A3,
Canada}
\author{A.P.C. Malbouisson}
\affiliation{Centro Brasileiro de Pesquisas F\'{\i}sicas, Rua Dr. X. Sigaud 150,
22290-180, Rio de Janeiro, RJ, Brazil}
\author{J.M.C. Malbouisson}
\affiliation{Instituto de F\'{\i}sica, Universidade Federal da Bahia,
40210-340, Salvador, BA, Brazil}

\author{A.E. Santana}
\affiliation{International Centre for Condensed Matter Physics Instituto de
F\'{\i}sica, Universidade de Bras\'{\i}lia, 70910-900, Bras\'{\i}lia, DF,
Brazil}

\begin{abstract}
Representations of the Poincar\'{e} symmetry are studied by using a Hilbert
space with a phase space content. The states are described by  wave functions (
quasi amplitudes of probability) associated with   Wigner functions (quasi
probability density). The gauge symmetry analysis provides a realization of the
Seiberg-Witten gauge theory for noncommutative fields.
\end{abstract}

\maketitle

\section{Introduction}
\label{sec:intro}

Considering the association of noncommutative geometry with string theory,
Seiberg and Witten\cite{seiberg} studied a string dynamics described by a
minimally coupled supersymmetric gauge field in a non-commutative space. The
result is a generalized gauge field $A^{\mu }$ with an antisymmetric tensor
field given by $F^{\mu \nu }=\partial _{\nu }A^{\mu }-\partial _{\mu }A^{\nu
}-i\{A^{\mu },A^{\nu }\}_{M}$, where $\{A^{\mu },A^{\nu }\}_{M}$ is the
Moyal-Poisson bracket. Here we present a realization of such a noncommutative
gauge theory, starting with representations of space-time symmetries in a phase
space manifold. In particular, we show that correlation functions are closely
associated with the Wigner function, describing bosons and fermions.

The idea of noncommutativity in $\mathbb{R}^{3}$ starts with a suggestion by
Heisenberg~\cite{heisenberg}, which was realized by Snyder~\cite%
{snyder,snyder2} and Yang~\cite{yang}, by exploring the de Sitter space. The
algebraic structure of non-commutative geometries has been developed
along different lines, in particular in association with c*-algebras~\cite%
{Connes}. The interest in noncommutative theories, since the 1990s, is due to a
variety of applications~\cite{szabo, seiberg}, including abelian and
non-abelian gauge theories~\cite{szabo,rivel4,marcel1,adolfo1}, gravity~\cite%
{kalau,kastler}, standard model for particles~\cite{varilly1,varilly2}, supersymmetry~\cite{rivel2} and in quantum Hall effect~%
\cite{belissard}. Issues such as the ultraviolet/infrared mixing and the
renormalizability of noncommutative theories have been also addressed~\cite%
{adolf1,adolf2,adolf3}.

The compatibility of non-commutative geometries and the Lorentz space-time
symmetry has been achieved with the twisted noncommutative theories~\cite%
{Balachandran:2006pi,Balachandran:2006hr,Balachandran:2011pd}. However, this
type of compatibility arises also in the Wigner-function formalism~\cite%
{wigner1,wigner2}, initially proposed for developments in quantum kinetic
theory. The algebraic structures and applications of Wigner-function are of
interest in a vast range of areas, including noncommutative geometries and
relativistic theories~\cite%
{wigner2,wig333,wig4,boo1,moy1,moy2,seb42,seb43,dayi,seb13,bos,seb445,dodo6,dodo8,berkow}%
. It is important to mention the use of this method in experiments considering
the reconstruction of quantum states and measurements of Wigner
function in quantum tomography~\cite%
{smithey,leibfried,davidovich,vogel,Ibort:2009bk}.

Associated with the algebraic structure of the Wigner formalism, there are
studies exploring the notion of wave function in phase-space~\cite%
{seb42,seb43,bos,seb445}, a crucial step to develop abelian and non-abelian
gauge theories, with a close connection to Wigner theory. However, several
aspects in these methods remain to be clarified; the most important being a
rigorous association of wave functions in phase-space (or the quasi-amplitude of
probability) with the Wigner function (the quasi distribution of probability),
in order to provide a physical interpretation for the
formalism~\cite{seb1,seb2,seb42,seb43,seb22}. This problem is addressed here in
order to carry out an analysis of a gauge theory associated with the Wigner
function. This leads to a realization of the Seiberg-Witten gauge theory for
noncommutative fields. We start, in Section 2, by presenting the
Poincar\'{e}-Lie algebra, using a carrier Hilbert space with the content of
phase space. This is used in Section 3 to derive spin zero and spin 1/2
representations. Then, in Section 4, the gauge theory is constructed for bosons
and fermions. Some final concluding remarks are presented in Section 5.

\section{Poncar\'{e} group in phase space}

Let $\mathbb{M}$ be an analytical manifold where each point is specified by
Minkowski coordinates $q^{\mu },$ with $\mu =0,1,2,3$, and a metric such that
$diag(g)\mathbf{=(+---)}$. Let $T^{\ast }\mathbb{M}$ be the cotangent-bundle,
where each point is specified by the coordinates $(q^{\mu },p^{\mu })$. We use
this twofold structure of $T^{\ast }\mathbb{M}$ to construct a standard
representation for the c$^{\ast }$-algebra with the content of a phase space.

We take advantage of the fact that $T^{\ast }\mathbb{M}$ can be equipped
with a symplectic structure via a 2-form $\omega =dq^{\mu }\wedge dp_{\mu }$%
, called the symplectic form. Let us define the operator on $C^{\infty }(T^{\ast
}M)$,
\begin{equation}
\Lambda =\frac{\overleftarrow{\partial }}{\partial q^{\mu }}\frac{
\overrightarrow{\partial }}{\partial p_{\mu }}-\frac{\overleftarrow{\partial
}}{\partial p^{\mu }}\frac{\overrightarrow{\partial }}{\partial q_{\mu }},
\label{fasenova2}
\end{equation}
such that for $C^{\infty }$ functions, $f(q,p)$ and $g(q,p),$ we have
\begin{equation}
\omega (f\Lambda ,g\Lambda )=f\Lambda g=\{f,g\},  \label{fasenova3}
\end{equation}
where
\begin{equation*}
\{f,g\}=\frac{\partial f}{\partial q^{\mu }}\frac{\partial g}{\partial p_{\mu
}}-\frac{\partial f}{\partial p^{\mu }}\frac{\partial g}{\partial q_{\mu }}
\end{equation*}
is the Poisson bracket. The space $T^{\ast }\mathbb{M}$ endowed with this
symplectic structure is called the phase space, and will be denoted by $%
\Gamma $.

In order to construct a Hilbert space in $C^{\infty }(\Gamma )$, let $%
\mathcal{H}(\Gamma )$ be a linear subspace of the space of measurable functions
$\psi :\Gamma \rightarrow \mathbb{C}$ which are square integrable, i.e. such
that
\begin{equation*}
\int_{\Gamma }d^{4}pd^{4}q\psi ^{\ast }(q,p)\psi (q,p)<\infty .
\end{equation*}%
The Hilbert space is introduced by defining the inner product, $\langle \cdot
|\cdot \rangle $, on $\mathcal{H}(\Gamma )$, as
\begin{equation*}
\langle \psi _{1}|\psi _{2}\rangle =\int_{\Gamma }\psi _{1}(q,p)^{\ast }\psi
_{2}(q,p)d^{4}pd^{4}q,
\end{equation*}%
where we take $(q^{\mu },p^{\mu })=(q,p)$, and $\psi (q,p)$ in $C^{\infty
}(\Gamma )$. In this case we have $\psi (q,p)=\langle q,p|\psi \rangle $, with
\begin{equation*}
\int d^{4}pd^{4}q|q,p\rangle \langle q,p|=1\ \ \mathrm{\ and}\ \ \langle
q,p\left\vert q^{\prime },p^{\prime }\right\rangle =\delta (q-q^{\prime })\delta
(p-p^{\prime }).
\end{equation*}%
Using the kets $|q,p\rangle $, we define the following c-number operators $%
\overline{Q}$ and $\overline{P}$ by
\begin{equation}
\overline{Q}|q,p\rangle =q|q,p\rangle ,\ \ \ \ \overline{P}|q,p\rangle
=p|q,p\rangle ,  \label{dav11}
\end{equation}%
fulfilling the commutation condition $[\overline{Q},\overline{P}]=0$. This
Hilbert space is taken here as the representation space of the Poincar\'{e}
symmetries. For the sake of physical interpretation, the state of a system will
be described by functions $\psi (q,p)$, with the normalization condition
\begin{equation}
\langle \psi |\psi \rangle =\int d^{4}pd^{4}q\psi ^{\ast }(q,p)\psi (q,p)=1
\end{equation}%
A unitary transformation in $\mathcal{H}(\Gamma )$ is the mapping $U:%
\mathcal{H}(\Gamma )\rightarrow \mathcal{H}(\Gamma )$ such that $\langle
\psi _{1}|\psi _{2}\rangle $ is invariant. We consider the mapping $%
U_{\Delta }=\exp (-i\hbar \Delta /2)$, where $\Delta =\frac{\partial }{%
\partial q^{\mu }}\frac{\partial }{\partial p_{\mu }}$. This linear
transformation gives rise to the following basic operators
\begin{eqnarray}
P^{\mu } &=&U_{\Delta }\overline{P}^{\mu }U_{\Delta }^{-1}=p^{\mu }-\frac{%
i\hbar }{2}\frac{\partial }{\partial q_{\mu }}, \label{jan20143} \\
Q^{\mu } &=&U_{\Delta }^{-1}\overline{Q}^{\mu }U_{\Delta }=q^{\mu }+\frac{%
i\hbar }{2}\frac{\partial }{\partial p_{\mu }}.\label{jan20144}
\end{eqnarray}%
In natural units ($\hbar =c=1$), which, unless explicitly stated, we use from
now on, these operators satisfy the following commutation relation:
\begin{equation*}
\lbrack Q^{\mu },P^{\nu }]=ig^{\mu \nu }.
\end{equation*}
Then we construct a representation for the Poincar\'{e} group in phase
space, by defining%
\begin{equation*}
M_{\mu \nu }=Q^{\mu }P^{\nu }-P^{\nu }Q^{\mu },
\end{equation*}%
such that the Poincar\'{e} Lie algebra is given by
\begin{align}
\lbrack M_{\mu \nu },P_{\sigma }]& =i(g_{\nu \sigma }P_{\mu }-g_{\sigma \mu
}P_{\nu }),  \label{poin1} \\
\lbrack M_{\sigma \rho },M_{\mu \nu }]& =i(g_{\mu \rho }M_{\nu \sigma
}-g_{\nu \rho }M_{\mu \sigma }  \notag \\
& +g_{\mu \sigma }M_{\rho \nu }-g_{\nu \sigma }M_{\rho \mu }); \label{poinc4}
\end{align}%
other commutation relations are zero. Explicitly, we have
\begin{align}
M^{\mu \nu }& =q^{\mu }p^{\nu }-q^{\nu }p^{\mu }-\frac{i}{2}q^{\mu }\frac{%
\partial }{\partial q_{\nu }}+\frac{i}{2}q^{\nu }\frac{\partial }{\partial
q_{\mu }}  \notag \\
& -\frac{i}{2}p^{\mu }\frac{\partial }{\partial p_{\nu }}+\frac{i}{2}p^{\nu }%
\frac{\partial }{\partial p_{\mu }}  \notag \\
& +\frac{i}{4}\frac{\partial ^{2}}{\partial p_{\mu }\partial q_{\nu }}-\frac{%
i}{4}\frac{\partial ^{2}}{\partial p_{\nu }\partial q_{\mu }}. \label{poinc44}
\end{align}

In order to construct representations, we use the Casimir invariants
\begin{equation*}
P^{2}=m^{2}\ \ \mathrm{and}\ \ w_{\mu }w^{\mu }=-m^{2}s(s+1),
\end{equation*}%
where
\begin{equation*}
w_{\mu }=\frac{1}{2}\varepsilon _{\mu \nu \rho \sigma }M^{\nu \sigma }P^{\rho
}=-m^{2}s(s+1)\
\end{equation*}%
is the Pauli-Lubanski vector.

\section{Scalar and Dirac fields in phase space}

For the scalar representation, we have $s=0$, such that $P^{\mu }P_{\mu }\phi
(p,q)=m^{2}\phi (p,q)$, that leads to
\begin{equation}
\frac{-1}{4}\frac{\partial ^{2}\phi (p,q)}{\partial q^{\mu }\partial q_{\mu }%
}-ip^{\mu }\frac{\partial \phi (p,q)}{\partial q^{\mu }}+(p^{\mu }p_{\mu
}-m^{2})\phi (p,q)=0.  \label{KGjan1}
\end{equation}%
This is a Klein-Gordon-like equation written in phase space. In order to
find the physical meaning for such an equation, we show the association of $%
\phi (p,q)$ with the Wigner function. Indeed it has been shown that~\cite%
{seb1,seb2}
\begin{equation}
f_{W}(q,p)=\phi (q,p)\star \phi ^{\dagger }(q,p),  \label{junho271}
\end{equation}%
satisfies the equation
\begin{equation}
\{p^{2},f_{W}(q,p)\}_{M}=p_{\mu }\frac{\partial \ }{\partial q^{\mu }}%
f_{W}(q,p)=0,  \label{junho272}
\end{equation}%
such that
\begin{equation*}
\{f(q,p),g(q,p)\}_{M}=f(q,p)\star g(q,p)-g(q,p)\star f(q,p)\
\end{equation*}%
\ is the Moyal bracket, where the star product given by
\begin{equation*}
A_{W}(q,p)e^{\frac{i\Lambda }{2}}B_{W}(q,p)=A_{W}(q,p)\star B_{W}(q,p).
\end{equation*}%
Thus we  introduce the Wigner mapping in the following way. Let $\widehat{A}$ be
an operator acting in the Hilbert space $\mathcal{H}$. The Wigner mapping $%
W:\widehat{A}\rightarrow A_{W}(q,p)$ is defined by
\begin{equation*}
A_{W}(q,p)=\frac{1}{\sqrt{2\pi }}\int dz\exp (ipz)\langle
q-\frac{z}{2}|\widehat{A}|q+%
\frac{z}{2}\rangle .
\end{equation*}%
Considering such a mapping for $f_{W}(q,p)$, with $W:\rho \rightarrow
f_{W}(q,p)$, then we show that~\cite{boo1}
\begin{equation*}
\lbrack \frac{\partial ^{2}\ }{\partial q^{\mu }\partial q_{\mu }},\rho ]=0.
\end{equation*}%
This is a Liouville--von Neumann-like equation. Since $\int dqdpf_{W}(q,p)=1$%
, we have Tr$\rho =1$, implying that $\rho $ is a density matrix and $%
f_{W}(q,p)$ is a Wigner function derived from $\rho $. In addition, it is
important to observe that if we consider, for instance,  $\rho (q)=\phi (q)\phi
^{\dagger }(q)$, with $\phi (q)\in \mathcal{H}$, then $\phi (q)$ satisfies the
Klein-Gordon equation,
\begin{equation*}
(\frac{\partial ^{2}}{\partial q^{\mu }\partial q_{\mu }}-m^{2})\phi (q)=0.
\end{equation*}

An operator $\widehat{A}$ acting on the Hilbert space $\mathcal{H}$ is mapped
into a Wigner representation by $W:A\rightarrow A_{W}(q,p)$, which
 in turn is mapped in
operators acting on $\mathcal{H}(\Gamma )$, such as
\begin{equation*}
\Omega :A_{W}(q,p)\rightarrow A(Q,P)=A_{W}(q,p)\star .
\end{equation*}%
Then we derive the composed mapping%
\begin{equation*}
\Omega \circ W:\widehat{A}\rightarrow A(Q,P).
\end{equation*}
As an example, consider, respectively, the position and momentum operators,
$\widehat{q}$ and $\widehat{p}$ defined by acting on the Hilbert space
$\mathcal{H}$, we have
\begin{equation*}
(\Omega \circ W)(\widehat{q}^{\mu})=Q^{\mu}=q^{\mu}\star ,\ \ \ (\Omega \circ W)
(\widehat{p}^{\mu}%
)=P^{\mu}=p^{\mu}\star
\end{equation*}%
where the operators $Q^{\mu}$ and $P^{\mu}$  were first introduced
 in Eq. (\ref{jan20143}) and Eq. (\ref{jan20144}), respectively.
  In addition, we prove the following
identity for the average of an observable $A(P,Q)$ in a state $|\phi \rangle \in
\mathcal{H}(\Gamma )$:
\begin{align*}
\langle A\rangle & =\langle \phi |A(P,Q)|\phi \rangle  \\
& =\int dqdp\phi ^{\dagger }(q,p)A_{W}(q,p)\star \phi (q,p) \\
& =\int dqdpA(q,p)f_{W}(q,p)=Tr\rho A,
\end{align*}%
where we have used$\ $%
\begin{equation*}
A(P,Q)=A_{W}(q,p)\star =A_{W}(q\star ,p\star ).
\end{equation*}%
It is important to emphasize that the association of $\phi (q,p)$ with the
Wigner function provides the physical interpretation of the formalism based on
representations in the Hilbert space $\mathcal{H}(\Gamma)$. In this case, since
the Wigner function is a quasi-distribution of probability, it is natural to
denominate $\phi (q,p)$ as a quasi-amplitude of probability.

Representations for fermions are carried out following standard procedures.
Taking $s=1/2$ in the Pauli-Lubanski vector, it leads to a Dirac equation in
phase space given by \cite{seb22}
\begin{equation}
\gamma ^{\mu }(p_{\mu }-\frac{i}{2}\frac{\partial }{\partial q^{\mu }})\psi
=m\psi ,  \label{asasa1}
\end{equation}%
where the $\gamma $-matrices satisfy $(\gamma ^{\mu }\gamma ^{\nu }+\gamma ^{\nu
}\gamma ^{\mu })=2g^{\mu \nu }$. The Lagrangian density is
\begin{equation}
\mathcal{L}=\frac{-i}{4}[(\partial \overline{_{\mu }\psi })\gamma ^{\mu }\psi
-\overline{\psi }(\gamma ^{\mu }\partial _{\mu }\psi )]-\overline{\psi
}(q,p)(m-\gamma ^{\mu }p_{\mu })\psi (q,p).  \label{asasa2}
\end{equation}%
In a similar manner as for the case of bosons, considered before, the Wigner
function associated with the Dirac field in phase space is
\begin{equation*}
f_{W}(q,p)=\psi (q,p)\star \overline{\psi }(q,p),
\end{equation*}
satisfying the equation of motion%
\begin{equation*}
p^{\mu }\frac{\partial f_{W}(q,p)}{\partial q^{\mu }}=0;
\end{equation*}%
such that, considering a mapping $W:\rho \rightarrow f_{W}(q,p)$,  a
Liouville--von Neumann equation is derived:
\begin{equation*}
\lbrack \gamma ^{\mu }\partial _{\mu },\rho ]=0.
\end{equation*}

It is worth mentioning that, due to the nature of the probability
quasi-amplitude, Eq.~(\ref{KGjan1}) and Eq.~(\ref{asasa1}) provide
phase-invariant field formalisms in phase space (which is not the case of the
usual Wigner function method). This aspect is developed in the following.

\section{Seiberg-Witten gauge fields}

The Lagrangian of the free Klein-Gordon field describing $N$-bosons is written
as
\begin{equation}
\mathcal{L}_{0}=(p^{\mu }\star \phi )(p_{\mu }\star \phi ^{\dagger })+m^{2}\phi
\phi ^{\dagger },  \label{la}
\end{equation}%
where $\phi \equiv \phi (q,p)$. From this lagrangian, Eq.~(\ref{KGjan1}) can be
derived. Our aim is to analyze the invariance of Eq.~(\ref{la}) under local
gauge transformations. In this sense, we take the transformation rules given by
\begin{equation}
\phi \rightarrow e^{-i\Lambda }\star \phi ,\ \ \ \ \phi ^{\dagger }\rightarrow
e^{i\Lambda }\star \phi ^{\dagger },  \label{t1}
\end{equation}%
and where $\Lambda \equiv \Lambda (q,p)$. For infinitesimal transformation,we
have $\delta \phi =-i\Lambda \star \phi \ $\ and $\delta \phi ^{\dagger
}=i\Lambda \star \phi ^{\dagger }$, such that
\begin{equation}
\delta (p^{\mu }\star \phi )=-ip^{\mu }\star \Lambda \star \phi ,  \label{t5}
\end{equation}%
and
\begin{equation}
\delta (p_{\mu }\star \phi ^{\dagger })=ip_{\mu }\star \Lambda \star \phi
^{\dagger }.  \label{t6}
\end{equation}%
It should be noted that, $(p^{\mu }\star \phi )$ and $(p_{\mu }\star \phi
^{\dagger })$ do not transform covariantly, i.e. not in the same way as $%
\phi $ and $\phi ^{\dagger }$ themselves. To demonstrate this aspect, we define
the operator
\begin{equation*}
D^{\mu }\star =p^{\mu }\star -iA^{\mu }\star ,
\end{equation*}
leading to a Lagrangian  written as
\begin{equation}
\mathcal{L}=(D^{\mu }\star \phi )(D_{\mu }\star \phi ^{\dagger })+m^{2}\phi \phi
^{\dagger }.  \label{la2}
\end{equation}%
Now we have to prove the invariance of this Lagrangian. It follows immediately
that
\begin{align}
\delta (D^{\mu }\star \phi )& =-ip^{\mu }(\Lambda \star \phi )-\frac{1}{2}%
\frac{\partial \Lambda }{\partial q_{\mu }}\star \phi   \notag \\
& -\frac{1}{2}\Lambda \star \frac{\partial \phi }{\partial q_{\mu }}-A^{\mu
}\star (\Lambda \star \phi )-i(\delta A^{\mu })\star \phi   \label{d1}
\end{align}%
Now, using the identity
\begin{equation*}
p(f\star g)=f\star (pg)-\frac{i}{2}(\partial _{q}f)\star g,
\end{equation*}
we have
\begin{align}
\delta (D^{\mu }\star \phi )& =-i\Lambda \star (p^{\mu }\star \phi )-\frac{%
\partial \Lambda }{\partial q_{\mu }}\star \phi   \notag \\
& -A^{\mu }\star (\Lambda \star \phi )-i(\delta A^{\mu })\star \phi . \label{d2}
\end{align}%
Considering a gauge transformations of the second kind, we demand that,
\begin{equation}
A^{\prime }{}^{\mu }\rightarrow A^{\mu }+i\{A^{\mu },\Lambda \}_{M}+i\frac{%
\partial \Lambda }{\partial q_{\mu }},  \label{moyal}
\end{equation}%
where $\{a,b\}_{M}=a\star b-b\star a$ is the Moyal bracket. Then Eq.~(\ref%
{d2}) reads
\begin{equation}
\delta (D^{\mu }\star \phi )=-i\Lambda \star (p^{\mu }\star -iA^{\mu }\star
)\phi ;  \notag
\end{equation}%
i.e.
\begin{equation}
\delta (D^{\mu }\star \phi )=-i\Lambda \star (D^{\mu }\star \phi ), \label{r1}
\end{equation}%
which is the covariant transformation rule.

Similarly, we show that
\begin{equation}
\delta (D_{\mu }\star \phi ^{\dagger })=-i\Lambda \star (D_{\mu }\star \phi
^{\dagger }).  \label{r2}
\end{equation}
Then we have the following rule for the minimal substitution: replace $%
p^{\mu }\star $ by $p^{\mu }\star -iA^{\mu }\star $.

The Lagrangian given by Eq.~(\ref{la2}) is now invariant. The field $A^{\mu }$ ,
however, must contribute by itself to the Lagrangian. Then we define
\begin{equation}
F^{\mu \nu }=\frac{\partial A^{\mu }}{\partial q_{\nu }}-\frac{\partial A^{\nu
}}{\partial q_{\mu }}-i\{A^{\mu },A^{\nu }\}_{M},
\end{equation}
so that $F^{\mu \nu }$ is invariant. Then the final Lagrangian is
\begin{equation*}
\mathcal{L}=(D^{\mu }\star \phi )(D_{\mu }\star \phi ^{\ast })+m^{2}\phi \phi
^{\ast }-\frac{1}{4}F^{\mu \nu }F_{\mu \nu }.
\end{equation*}
This provides a realization of the Seiberg-Witten gauge theory for
noncommutative fields.

Proceeding with  the usual canonical quantization, we take $\phi $ as an
operator acting in a  (symplectic) Fock  space, such that the two-point function
for the free field is defined as
\begin{equation}
G(q-q^{\prime },p-p^{\prime })=\langle 0|\mathrm{T}[\phi (q,p)\phi ^{\ast
}(q^{\prime },p^{\prime })]|0\rangle ,  \label{twopoint}
\end{equation}%
where T is the time-ordering operator. The physical interpretation of $%
G(q-q^{\prime },p-p^{\prime })$ is obtained by observing that it is related to
the Wigner function by~\cite{seb22}
\begin{equation}
f_{W}(q,p)=\lim_{q^{\prime },p^{\prime }\rightarrow q,p}\exp {i(\partial
_{q}\partial _{p^{\prime }}-\partial _{q^{\prime }}\partial _{p})}%
G(q-q^{\prime },p-p^{\prime }).  \label{tpw}
\end{equation}%
This provides a way to develop a perturbative theory for the Wigner function
considering this scalar electrodynamics in phase space. Observe that we have the
mapping: $\Omega \circ W:\mathrm{T}[\phi (q)\phi ^{\ast }(q^{\prime
})]\rightarrow \mathrm{T}[\phi (q,p)\phi ^{\ast }(q^{\prime },p^{\prime })]$%
, such that  $\langle 0|\mathrm{T}[\phi (q)\phi ^{\ast }(q^{\prime })]|0\rangle
$ is the usual propagator for the free Klein-Gordon field. For the gauge field,
considering the changing  $\phi \rightarrow A^{\mu }$, we obtain similar
correlation functions, with the same physical interpretation.

Considering then the gauge invariance, we obtain a realization of the
Seiberg-Witten gauge theory for fermions
\begin{equation*}
\mathcal{L}=\ \overline{\psi }(q,p)(\gamma ^{\mu }D_{\mu }\star \psi
(q,p)+m^{2}\overline{\psi }(q,p)\psi (q,p)-\frac{1}{4}F^{\mu \nu }F_{\mu \nu }.
\end{equation*}
This gauge invariant Lagrangian in phase space provides a theory associated with
a Wigner function for spin 1/2 particles.

\section{Concluding remarks}

In this paper, we have derived the field theory in phase space, associated with
Wigner functions. Taking into account the notion of wave-functions in
phase-space (quasi-amplitudes of probability), we are able to introduce gauge
fields in phase space. Then equations for the quantum electrodynamics in phase
space are derived. It is important to emphasize that the gauge symmetry provides
a realization of the Seiberg-Witten gauge theory for noncommutative fields. The
generalization for non-abelian gauge fields, although intricate, follows along
similar lines. This aspect will be analyzed closely elsewhere.

\acknowledgments

This work was partially supported by CAPES and CNPq of Brazil and NSERC of
Canada.

\end{document}